\documentclass[pageno]{jpaper}

\usepackage[normalem]{ulem}

\usepackage{amsmath}
\usepackage{times}
\usepackage{makeidx}  % allows for indexgeneration
\usepackage{titling}

\usepackage{listings}
\usepackage{inconsolata}

\usepackage{multirow}
\usepackage{setspace}
\usepackage[affil-sl]{authblk}
\usepackage[font=small, skip=1pt]{caption}
\usepackage{amssymb}
\usepackage[compact]{titlesec}
\usepackage{array}
\usepackage{ctable}
\usepackage{float}
\usepackage{graphicx}
\usepackage{chngcntr}
\usepackage{mathptmx}

\usepackage[numbers]{natbib}
\usepackage{subfig}
\usepackage{balance}

\usepackage{backnaur}

\newcommand{\sys}{\emph{Espresso}}

\lstdefinelanguage{java}{
  morekeywords={abstract,case,catch,class,def,%
    do,else,extends,false,final,finally,%
    for,if,implicit,import,match,mixin,%
    new,null,object,override,package, public,%
    private,protected,requires,return,sealed,%
    super,this,throw,trait,true,try,%
    type,int,double,while,pnew, persistable},
  otherkeywords={=>,<-,<\%,<:,>:,\#,@},
  sensitive=true,
  morecomment=[l]{//},
  morecomment=[n]{/*}{*/},
  morestring=[b]",
  morestring=[b]',
  morestring=[b]""",
}

\lstnewenvironment{q1}[1][]
{\lstset{
frame=none,
language=java,
aboveskip=3mm,
belowskip=3mm,
showstringspaces=false,
columns=flexible,
breaklines=true,
basicstyle={\footnotesize\ttfamily}, 
tabsize=4,  
numbers=left,
xleftmargin=2em,
framexleftmargin=2em,
#1} }
 {}
 
\newcommand{\subtitle}[1]{%
  \posttitle{%
    \par\end{center}
    \begin{center}\large#1\end{center}
      \vskip0.5em}%
  }

\begin{document}

\title{Espresso:  Brewing Java For More Non-Volatility with Non-volatile Memory}
%\subtitle{A Case for Scalable and Efficient Middleboxs with Container-based Virtualization}

\author{\rm Mingyu Wu, Ziming Zhao, Haoyu Li, Heting Li, Haibo Chen, Binyu Zang, Haibing Guan\\
  Shanghai Key Laboratory of Scalable Computing and Systems  \\
  Institute of Parallel and Distributed Systems, Shanghai Jiao Tong University}
 
 \date{}
            
\maketitle

\pagestyle{empty}

\begin{abstract}

Fast, byte-addressable non-volatile memory (NVM) embraces both near-DRAM latency and disk-like persistence,
which has generated considerable interests to revolutionize system software stack and programming models.
However, it is less understood how NVM can be combined with managed runtime like Java virtual machine (JVM)
to ease persistence management. 
This paper proposes \emph{Espresso\footnote{Espresso coffee contains 
more non-volatile chemicals (such as caffeine); we use it as an analog to our work where data becomes more non-volatile}}, 
a holistic extension to Java and its runtime, to enable Java programmers to exploit NVM for persistence management with high performance. 
Espresso first provides a general persistent heap design called Persistent Java Heap (PJH) to manage persistent data as normal Java objects. 
The heap is then strengthened with a recoverable mechanism to provide crash consistency for heap metadata. 
It then provides a new abstraction called Persistent Java Object (PJO) to provide an easy-to-use but safe persistent programming model for 
programmers to persist application data.  
The evaluation confirms that \emph{Espresso} significantly outperforms state-of-art NVM support for 
Java (i.e., JPA and PCJ) while being compatible to existing data structures in Java programs.

\end{abstract}

\section{Introduction}
\label{sec:intro}

Due to promising features like non-volatility, byte-addressability and close-to-DRAM speed, emerging non-volatile
memories (NVM) are projected to revolutionize the memory hierarchy in the near future. In fact, battery-backed
non-volatile DIMM (NVDIMM)~\cite{nvdimm} has been available to the market for years. With the official release 
of Intel and Micron's 3D-Xpoint~\cite{3dxpoint} to the market, it is foreseeable to see NVM to be widely deployed soon. 

While there have been considerable interests to leverage NVM to boost the performance and ease the persistence management
of native code~\cite{coburn2011nv, hsu2017nvthreads, NVML, liu2017dudetm, memaripour2017atomic, nalli2017analysis, pelley2014memory, volos2011mnemosyne}, 
how NVM can be exploited by high-level programming languages with managed runtime like Java is less 
understood. Despite their attracting features such as automatic memory management, portability and productivity, the additional 
layer of abstraction brought by the language virtual machine (e.g., JVM) complicates the persistence management. 

The mainstream persistent programming model leverages a coarse-grained abstraction like Java Persistence API (JPA)~\cite{demichiel2006java} 
to provide easy-to-use transactional APIs for programmers to persist their data.  However, it does not consider the emergence of NVM, 
and creates unnecessary transformation overhead between Java objects and native serialized data.  In contrast, the recent proposed 
Persistent Collections for Java (PCJ)~\cite{PCJ} provides a fine-grained programming model to enable users to manipulate persistent data in object level. 
However, it has built an independent type system against the original one in Java, which makes it hard to be compatible with existing Java programs since
it mandates the use of the collections defined by itself. Furthermore, PCJ manages persistent data as native objects on their own,
which ends up with poor performance\footnote{The home page (https://github.com/pmem/pcj) of PCJ acknowledged that ``The breadth of persistent types is currently limited and the code is not performance-optimized''.}.  Besides, these two approaches target different application scenarios and programmers cannot uniformly 
use one approach to applications that have both requirements. 

This paper proposes {\sys}, a unified persistence framework that supports both fine-grained and coarse-grained 
persistence management while being mostly compatible with data structures in existing Java programs and notably 
boost the persistence management performance. {\sys} provides \emph{Persistent Java Heap (PJH)}, an NVM-based heap to seamlessly 
store persistent Java objects. PJH allows users to manipulate persistent objects as if they were stored in a normal Java heap
and thus requires no data structure changes. To allocate data on PJH, {\sys} provides a lightweight keyword \emph{pnew} 
to create Java objects in NVM. 

PJH serves as an NVM-aware allocator, which should tolerate machine crashes to create a safe runtime environment for programmers. 
Hence, PJH provides crash-consistent allocation and deallocation (garbage collection), which guarantee that the metadata 
of the heap is crash consistent and robust against failures. 

To further ease the programming for applications that require coarse-grained persistence, {\sys} provides the Persistent Java Object (PJO), 
a new persistent programming abstraction atop PJH as a replacement of JPA for NVM.  PJO provides backward-compatibility by reusing
the annotations and transactional APIs in JPA, yet with additional optimizations to eliminate unnecessary overhead 
in original JPA to boost the performance. 

We have implemented the design of PJH and PJO atop OpenJDK 8.0. To confirm the effectiveness of our design, we provide a set of evaluation 
against JPA and PCJ.  
The result indicates that {\sys} achieves up to 256.3x speedup compared with PCJ for a set of microbenchmarks. Furthermore, PJO can 
provide support for different kind of data types in the JPAB benchmark, which gains up to 3.24x speedup over the original JPA for the H2 
database. 

In summary, this paper makes the following contributions:
\begin{itemize}
\item A persistent Java heap design (PJH) that enables Java programs to exploit NVM for persistence management without massive reengineering. 
\item A new abstraction for persistent programming (PJO) for simple and safe manipulation on persistent data objects. 
\item An implementation of PJH and PJO atop OpenJDK and a set of evaluations to confirm its effectiveness.
\end{itemize} 

The rest of our paper is organized as follows. 
Section~\ref{sec:background} reviews two main approaches for persistence management and discusses its deficiencies, which motivates the design of {\sys}. 
Section~\ref{sec:heap} introduces the overview of our PJH and language extension to manipulate persistent data objects. 
Section~\ref{sec:crash} further describes our mechanism to guarantee the crash consistency of PJH. 
Section~\ref{sec:jpo} presents the abstraction PJO together with an easy-to-use persistent programming model for programmers who require safe ACID semantics. 
We evaluate our design in section~\ref{sec:eval}, discuss related work in section~\ref{sec:related} and finally conclude in section~\ref{sec:concl}.

\section{Background and Motivation}
\label{sec:background}

In this section, we briefly review two main approaches for persistence management in Java, which provide coarse-grained and fine-grained persistence accordingly.  
We show that both approaches have some deficiencies in providing a compatible and efficient way for persistence.

\subsection{Coarse-grained Persistence with JPA}
\label{coarse-grained}

Database is a very appealing application for NVM and has been intensively studied by prior work~\cite{nalli2017analysis, seo2017failure, volos2011mnemosyne, yang2015nv}. 
Many databases~\cite{java-dbs} are written in Java due to the portability and easy programming. For ease of persistent programming, such databases usually provide 
a \emph{persistent layer} to keep programmers away from the messy work on persistent data management. 
This layer can be implemented according to Java official specification or a customized one for different use cases. Overall, it mainly serves data transformation between 
Java runtime and persistent data storage. 

A well-known example for the persistent layer is \emph{Java Persistent API} (JPA)~\cite{demichiel2006java}, which is a specification officially offered by the 
Java community for persistence programming, especially in relational database management system (RDBMS). 
With JPA, programmers are allowed to declare their own classes, sub-classes and even collections with some annotations. 
JPA is responsible for data transformation between Java applications and RDBMSes: it serves applications with objects while it communicates with RDBMSes 
via the Java Database Connectivity (JDBC) interface. JPA also provides the abstraction of ACID transactions for programmers,
which guarantees that all updates related to persistent data will be persisted after a transaction commits.

\begin{figure}[h]
\begin{minipage}{.99\linewidth}
\centering
\includegraphics[scale=0.35]{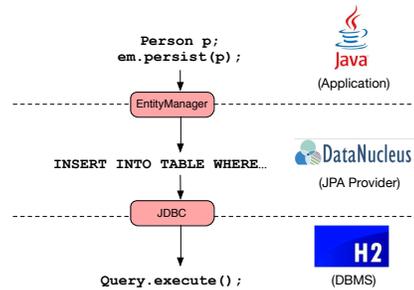}
\caption{The infrastructure of DataNucleus}
\label{fig:datanucleus}
\end{minipage} \\[-7pt]
\end{figure}

To understand the performance of JPA atop NVM, we present a case study with DataNucleus~\cite{DataNucleus}, a widely-used open-source implementation of JPA.
The architecture is shown in Figure~\ref{fig:datanucleus}. 

DataNucleus requires all classes related to persistent data to implement the \emph{Persistable} interface. 
Programmers should mark their classes with the annotation \emph{@persistable}. 
Suppose a programmer wants to declare a simple class \emph{Person} which contains two fields: \emph{id} (Integer) and \emph{name} (String), 
she should write code similar to that shown in Figure~\ref{fig:person-jpo}. 
Note that we will use the \emph{Person} class throughout the paper as a running example.  
DataNucleus has provided a bytecode instrumentor named \emph{enhancer} to transparently transform arbitrary classes with \emph{@persistable}  
into those with \emph{Persistable} interface implemented.  
Afterwards, the implementation of APIs required by \emph{Persistable} interface will also be automatically generated by the \emph{enhancer}. 
The enhancer will also insert some \emph{control fields} (corresponding to \emph{data fields} that store useful user data) into \emph{Persistable} objects and 
instrument user-defined methods (\emph{getId} in this example) for the ease of management.

\begin{figure}[h]
\begin{minipage}{.99\linewidth}
\centering
\begin{q1}
@persistable
public class Person {
	// fields
	private Integer id;
	private String name;
	
	// constructor
	public Person(Integer id, String name) {
		this.id = id;
		this.name = name;
	} 
	
	// a method example
	public Integer getId() {
		return this.id;
	}
	
	......
}
\end{q1}
\end{minipage} \\[-10pt]
\caption{The declaration for class \emph{Person} under JPA}
\label{fig:person-jpo}
\end{figure}

\vspace{2mm}

In DataNucleus, objects backed by persistent storage are managed by \emph{EntityManager} (em). 
EntityManager is also responsible for transaction management. 
As illustrated in Figure~\ref{fig:person-jpa}, programmers who want to persist their data in an ACID fashion can firstly initiate a transaction. 
Afterward, they can invoke \emph{em.persist} on the newly created object \emph{p};
EntityManger will add \emph{p} to its management list for future manipulation. 
Each managed object will also be associated with a \emph{StateManager} for state management. 
The reference to \emph{StateManager} is inserted into Persistable objects by the enhancer. 

\begin{figure}[h]
\begin{minipage}{.99\linewidth}
\centering
\begin{q1}
// Start a transaction
em.getTransaction().begin();
Person p = new Person(...);
em.persist(p);	
// Transaction commits
em.getTransaction().commit();
\end{q1}
\end{minipage} \\[-10pt]
\caption{Programming in JPA with ACID semantic}
\label{fig:person-jpa}
\end{figure}

The real persistence work happens when \emph{commit} is invoked. 
DataNucleus will find all modified (including newly added) objects from its management list and translate all updates into SQL statements. 
It subsequently sends the statements to the RDBMSes through JDBC to update data in the persistent storage. 
Note that only SQL statements are conveyed to DBMSes. 
Hence, even the RDBMS written in pure Java (like H2~\cite{mueller2012h2} in Figure~\ref{fig:datanucleus}) can only update databases with 
SQL instead of real data stored in objects. 

\begin{figure}[htbp]
\begin{minipage}{.99\linewidth}
\centering
\includegraphics[scale=0.4]{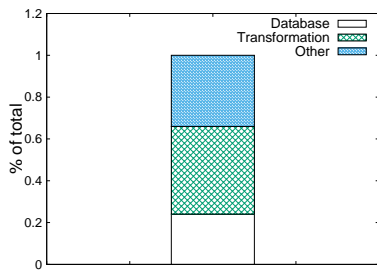}
\caption{Breakdown for commit phase of DataNucleus}
\label{fig:breakdown}
\end{minipage} \\[-10pt]
\end{figure}

\textbf{Deficiencies of JPA on NVM.} Yet, the data transformation phase in JPA induces significant overhead in the overall execution. 
We test its retrieve operation using the JPA Performance Benchmark (JPAB)~\cite{JPAB}.
Figure~\ref{fig:breakdown} illustrates the breakdown of performance. 
Surprisingly, the user-oriented operations on the database only account for 24.0\%. 
In contrast, the transformation from objects to SQL statements takes 41.9\%. 
This indicates that the JPA incurs notable performance overhead.
 
\subsection{Fine-grained Persistence with PCJ} 
\label{subsec:pcj}

To our knowledge, Persistent Collections for Java (PCJ~\cite{PCJ}) by Intel is the only active project to allow Java programmers to store their data in NVM. 
However, our study shows that PCJ also has several deficiencies due to its design.

\textbf{Separated type system.} 
PCJ implements a new type system based on a persistent type called \emph{PersistentObject}, and only objects whose type is a subtype of \emph{PersistentObject} 
can be stored in NVM. 
Users who want to use PCJ must extend \emph{PersistentObject} to implement their own types.  
Figure~\ref{fig:person-pcj} illustrates the declaration of the \emph{Person} class on PCJ\footnote{The original declaration of \emph{Person} is much more complex with a bunch of static variables and helper methods. We have simplified the declaration for ease of understanding.}. 
The class \emph{Person} must first extend \emph{PersistentObject} to fit PCJ. 
Furthermore, the type of \emph{id} and \emph{name} should be modified into \emph{PersistentInteger} and \emph{PersistentString} respectively, 
both of which are subtypes of \emph{PersistentObject}. Hence, using PCJ mandates a non-trivial reengineering to transform existing 
data structures to the form of those supported by PCJ.

\begin{figure}[h]
\begin{minipage}{.99\linewidth}
\centering
\begin{q1}
public class Person extends PersistentObject {
	// fields
	private PersistentInteger id;
	private PersistentString name;
	
	// constructor
	public Person(Integer id, String name) {
		this.id = new PersistentInteger(id.intValue());
		this.name = new PersistentString(name);
	} 
	
	// a method example
	public Integer getId() {
		return this.id.intValue();
	}
	
	......
}
\end{q1}
\end{minipage} \\[-10pt]
\caption{The declaration for a simple class \emph{Person} in PCJ}
\label{fig:person-pcj}
\end{figure}

\begin{figure}[h]
\begin{center}
\includegraphics[scale=.35]{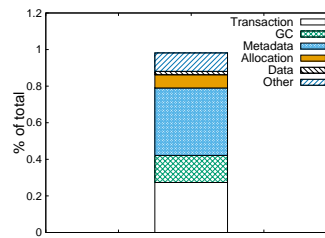}
\caption{Breakdown analysis for create operations in PCJ}
\label{fig:breakdown-pcj}
\end{center}
\end{figure}

\textbf{Off-heap data management. }
Due to the lack of support from Java, PCJ stores persistent data as native off-heap objects and manage them with the help of NVML~\cite{NVML}, 
a C library providing ACID semantics for accessing data in NVM. 
Therefore, PCJ has to define a special layout for native objects and handle synchronization and garbage collection all by itself. 
This may lead to non-trivial management overhead and suboptimal performance.  
We have implemented a simple example where we create 200,000 \emph{PersistentLong} objects (the equivalent of Java \emph{Long} in PCJ) 
and analyzed its performance in Figure~\ref{fig:breakdown-pcj}. 

First, the operation related to real data manipulation only accounts for 1.8\% over the whole execution time. 
In contrast, operations related to metadata update contributes 36.8\%, most of which is caused by type information memorization. 
In a normal Java heap, the type information operation only contains a pointer store, which is much simpler. 

Furthermore, it takes 14.8\% of the overall time to add garbage collection related information to the newly created object. 
PCJ needs this step because it is based on a reference counting GC algorithm, which needs to update GC-related information for each initialization. 
A normal Java heap leverages more mature garbage collectors and takes less time to bookkeep objects. 

The last source of overhead comes from transactions, which mainly contain synchronization primitives and logging.  
This phase can also be optimized with the reserved bit in object header and transaction libraries written in Java, if the objects are managed within Java heap. 

In summary, most overhead in PCJ is caused by its off-heap design, which could be notably optimized using an on-heap design. 

\subsection{Requirements for Persistence Management in Java}

From our study, we can see that there is currently no unified framework to provide persistence in Java. JPA is mostly useful for databases that require coarse-grained persistence, 
while PCJ mandates the use of the defined collections in order to enjoy fine-grained persistence, which would incur non-trivial porting efforts due to a shift of data structure. Besides,
both suffer from notable performance overhead and thus cannot fully exploit the performance benefit of NVM.

In light of this, we believe an ideal persistent framework for Java should satisfy the following requirements.

\begin{itemize}
\item \emph{Unified persistence:} The framework should support both fine-grained and coarse-grained persistence so as to support a wide range of applications. 
\item \emph{High performance:}  The framework should fully incur only a small amount of overhead for persistence to harness the performance advantage of NVM.
\item  \emph{Backward compatible:} The framework should not require major database changes so that existing applications can be ported with small effort 
to run atop it.
\end{itemize}

\section{Persistent Java Heap}
\label{sec:heap}

Being aware of the requirements described above, \emph{Espresso} uses a unified persistence management framework for 
Java to support both fine-grained and coarse-grained persistence management. It mainly contains two parts: Persistent Java 
Heap (PJH) to manage persistent objects in a fine-grained way, while Persistent Java Object (PJO) helps programmers to manage 
persistent data with handy interfaces. This section will mainly describe the design of PJH. 

\subsection{Overview}
\label{subsec:pjh-overview}

\begin{figure}[h]
\begin{minipage}{.99\linewidth}
\centering
\includegraphics[scale=0.5]{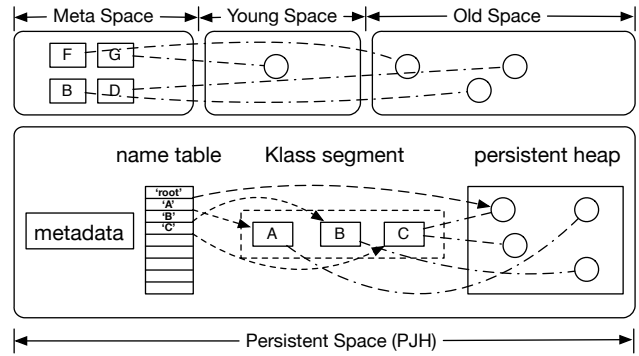}
\caption{The Java heap layout with PJH}
\label{fig:NVMheap}
\end{minipage} \\[-6pt]
\end{figure}

PJH is an extension to the original Java heap by adding an additional persistent heap.  We have built PJH in the \emph{Parallel Scavenge Heap} (PSHeap), the default implementation for Java heap.  Figure~\ref{fig:NVMheap} illustrates the modified layout of PSHeap with PJH. 
 
The original implementation of PSHeap contains two different spaces to store objects (circles in Figure~\ref{fig:NVMheap}): Young Space and Old Space. 
Objects will be initially created at the Young Space and later promoted to the Old Space if they have survived several collections. 
The garbage collector, namely Parallel Scavenge Garbage Collector (PSGC), also provides two different garbage collection algorithms. 
Young GC only collects the garbage within the Young Space, which happens frequently and finishes soon. 
In contrast, Old GC collects the whole heap, which is slow and thus happens infrequently. 

In Java, each object should hold a class pointer to its class-related metadata, which is called a \emph{Klass} in OpenJDK (rectangles in Figure~\ref{fig:NVMheap}). 
The class pointer is stored in the header of an object, right next to the real data fields (dashed lines in Figure~\ref{fig:NVMheap}).
JVM has maintained a \emph{Meta Space} to manage the Klasses. 
Klasses are very important because they store the layout information for objects. 
If the class pointer in an object is corrupted, or the metadata in Klass is lost, the data within the object will become uninterpretable. 

PJH is implemented as an independent \emph{Persistent Space} against the original PSHeap. 
It is arranged as a non-generational heap, and the garbage collection algorithm resembles the old GC in PSGC in that it is designed 
for long-lived objects and infrequent collections. The main components of it includes metadata area, name table, Klass segment and data heap. 
All the components should be persisted in NVM to guarantee the availability of the PJH.

\textbf{Data heap and Klass segment. }
Java objects required to be persisted are stored in the \emph{data heap}. 
The object header layout is the same as one in the normal Java heap, so each persistent object still holds a class pointer to its class-related metadata to its Klass. 
All Klasses used by persistent objects are stored in the \emph{Klass segment} and managed separately from the original Meta Space.  
%The Klasses are arranged as linked lists and managed together with other volatile Klasses. 

\textbf{Name table. }
The name table stores mappings from string constants to two different kinds of entries: Klass entries and root entries. 
A Klass entry stores the start address of a Klass in the Klass segment, which is set by JVM when an object is created in NVM while its Klass does not exist in the Klass segment. 
A root entry stores the address of a \emph{root object}, which should be set and managed by users. 
Root objects are essential especially after a system reboot, since they are the only known entry points to access the objects in data heap.  

\begin{figure}[h]
\begin{center}
\vspace{0.01\textwidth}
\includegraphics[scale=0.4]{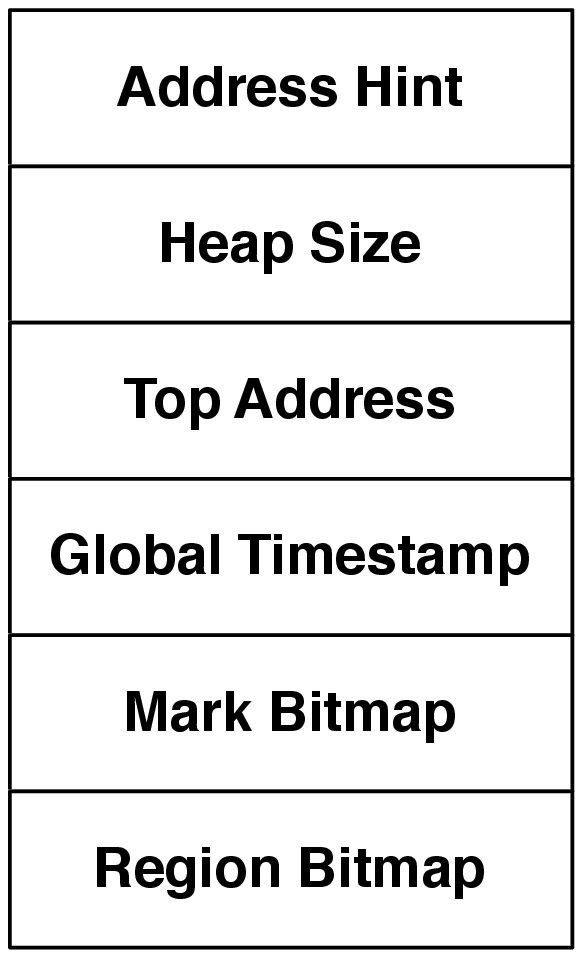}
\caption{The components for the metadata area in PJH}
\label{fig:meta}
\end{center}
\end{figure}

\textbf{Metadata area. }
The metadata area shown in Figure~\ref{fig:meta} is kept for memorizing heap-related metadata to build a reusable and crash-proof heap. 
The \emph{address hint} stores the starting virtual address of the whole heap for future heap reloading, while the \emph{heap size} 
stores the maximum address space the PJH can occupy. 
The \emph{top address} can be used to calculate the allocated bytes of PJH.
Other information is essential to implement a recoverable garbage collector for PJH and will be discussed in detail later.   

%The root objects are essential especially when users want to access persistent objects after a system reboot, as they are the only known 

%We also provide a name table to store strings for the known \emph{names}, i.e. the name of known root objects (served as entry points to the persistent heap) and Klasses. 
%Finally, the metadata area is kept for memorizing heap-related metadata, whose components are shown in Figure~\ref{fig:meta}. 
%The \emph{address hint} stores the starting virtual address of the whole heap for future heap reloading, while the \emph{heap size} stores the maximum address space the PJH can occupy. 
%The \emph{top address} can be used to calculate the allocated bytes of PJH.
%Other components are essential to implement a recoverable garbage collector for PJH and will be discussed in detail later.   

\subsection{Language Extension: pnew}
 
To allow users to create objects on NVM, we add a keyword \emph{pnew} to the Java programming language. 
The keyword has similar syntax rules to \emph{new} except that the corresponding objects will be laid on NVM. 
We have modified Javac to convert \emph{pnew} into four different bytecodes accounting for different syntaxes: \emph{pnew} (normal instances), \emph{panewarray} (object arrays), \emph{pnewarray} (primitive arrays) and \emph{pmultianewarray} (multi-dimensional object arrays). 
Those bytecodes will put objects into PJH regardless of their types.
Note that the keyword \emph{pnew} only allocates an object on NVM without considering its fields. 
If users want to make certain fields of an object persistent, they may need to implement a new constructor with \emph{pnew}. 

The keyword \emph{pnew} enables programmers to tackle with NVM in a very familiar way. 
Figure~\ref{fig:person-pjh} shows how to define the class \emph{Person} in section~\ref{sec:background}. 
Since it does not impose restrictions on which type can be persisted, the class \emph{Person} does not need to be extended from any particular types, 
nor do its fields need to be changed. The resulting code is very similar to that written for original Java except for the \emph{pnew} keyword and 
the newly added constructor for \emph{String}, so it is easy to understand.  

Note that the \emph{pnew} keywords in the constructor can be freely replaced with \emph{new}, as we do not force the invariants for references at the language level. 
Users are allowed to define pointers to volatile memory to support applications using a mix of NVM and DRAM. 
However, this design must take memory safety into consideration to avoid undefined memory behavior.  

\begin{figure}[h]
{
\begin{q1}
public class Person {
	// fields
	private Integer id;
	private String name;
	
	// constructor
	public Person(Integer id, String name) {
		this.id = pnew Integer(id);
		this.name = pnew String(name, true);
	} 
	
	// a method example
	public Integer getId() {
		return this.id;
	}
	
	......
}
\end{q1}
}
\caption{The declaration for class \emph{Person} atop PJH}
\label{fig:person-pjh}
\end{figure}

\textbf{Alias Klasses:} Our design allows objects of the same type to be stored in both DRAM and NVM, which violates the assumption 
of original Java runtime and raises some challenge. 
In Java, each Klass will contain a data structure called \emph{constant pool}~\cite{lindholm2014java}. 
Constant pools store important symbols which will be resolved during runtime. 
For each class symbol, a constant pool will initially create a slot and store a reference to its name (a string constant). 
After symbol resolution, the slot will instead store the address of the corresponding Klass. 

This design works perfectly in the stock JVM, but it induces some problem in PJH.
Consider the code in Figure~\ref{fig:simple} where we subsequently create two objects \emph{a} and \emph{b} of type \emph{Person} with \emph{new} and \emph{pnew} respectively. 
Afterwards, code in line 3 tries to cast the object type into \emph{Person}, which should have been a redundant type casting operation. 
Nevertheless, the program ends up with a \emph{ClassCastException}. 

\begin{figure}[t]
{
\begin{q1}
Person a = new Person(...);
Person b = pnew Person(...);
somefunc((Person)a);
// ClassCastException here!
\end{q1}
}
\caption{A simple program encountering wrong exception when using \emph{pnew}}
\label{fig:simple}
\end{figure} 

The problem happens because \emph{Person} objects are stored in both volatile and non-volatile memory, resulting in two different Klasses. 
Meanwhile, the constant pool keeps only one slot for each class symbol. 
In the example, 
JVM will find that the object is volatile and allocate the corresponding Klass for \emph{Person} (denoted as $K_{p}$) in DRAM when creating \emph{a}. 
Afterwards, JVM soon realizes that object \emph{b} should be persistent, so it also creates a Klass for \emph{Person} again in the Klass Segment in PJH ($K'_{p}$). 
Since the addresses for two Klasses differ, the constant pool has to store the address of $K'_{p}$ to replace that of $K_{p}$. 
On type casting, JVM finds that the resolved class in its constant pool ($K'_{p}$) is at odds with the type of \emph{a} ($K_{p}$), so it throws an exception. 

We introduce a concept named \emph{alias Klass} to handle this problem. 
Two Klasses are an alias to each other if they are logically the same class but stored in different places (NVM and DRAM). 
We add the alias check into type checking within JVM to avoid wrong exceptions. 
We have also extended the type lattice in the OpenJDK Server Compiler~\cite{paleczny2001java} to consider alias. 
Original type-related checks like~\cite{liang1998dynamic} are also extended.

\subsection{Heap Management}

In our programming model, users are allowed to create multiple PJH instances served for various applications. 
They are also required to define root objects as handles to access the persistent objects even after a system reboot. 
We have implemented some basic APIs in Java standard library (JDK) (shown in Table~\ref{tab:apis}) to help them manage the heap instances and root objects. 
%Users are allowed to create PJH instances, load existing instances and manipulate root objects through those APIs. 
Those APIs can be classified into two groups: \emph{createHeap}, \emph{loadHeap} and \emph{existsHeap} are 
heap-related while \emph{setRoot} and \emph{getRoot} are root-related. 
Figure~\ref{fig:management-apis} shows a simple example where we want to locate some data in a heap or initialize the heap if it does not exist. 

%\vspace{5mm}

\begin{table}[h]
\fontsize{7.5}{9.6}\selectfont
\centering
\begin{tabular}{lll}
	\hline
	API & Args & Description \\
	\hline
	createHeap & {name, size} & create a PJH instance\\
	
	loadHeap & {name} & load a PJH instance into current JVM\\
	
	existsHeap & {name} & check if a PJH instance exists\\
	
	setRoot & {name, object} & mark an object as a root\\
	
	getRoot & {name} & fetch a root object\\
	\hline
\end{tabular}
\vspace{3mm}
\caption{APIs for PJH management}
\label{tab:apis}
\end{table}

\vspace{5mm}

\begin{figure}[h]
{
\begin{q1}
// Check if the heap exists
if (existsHeap("Jimmy")) {
	// If so, load the heap and fetch objects
	loadHeap("Jimmy");
	Person p = (Person) getRoot("Jimmy_info"); 
} else {
	// Otherwise, create new heap and objects
	long size = 1024 * 1024;
	createHeap("Jimmy", size);
	Person p = pnew Person(...);
	setRoot("Jimmy_info", p);
}
\end{q1}
}
\caption{A simple example using heap management APIs}
\label{fig:management-apis}
\end{figure}

%\vspace{2mm}

\textbf{Heap-related APIs.} 
Java programmers can invoke \emph{createHeap} (line 9) to create a PJH instance with specified name and size (in bytes).  
%Since we simulate the NVRAM by using kernel \emph{memmap} parameters and DAX mapped files, we maintain a mapping table between name and PJH starting (virtual) address at the head of the NVRAM-mapped file.  
We have implemented an external name manager responsible for the mapping between the real data of PJH instances and their names. 
\emph{createHeap} will notify the name manager to insert a new mapping into the table. 
Furthermore, the starting (virtual) address should also be stored as \emph{address hint} in the metadata area of the PJH instance for future use. 
Afterwards, users can use \emph{pnew} to allocate objects on the newly created heap (line 10). 

Users are allowed to load pre-existing PJH instances into current JVM by invoking \emph{loadHeap}. 
They can optionally call \emph{existsHeap} in advance (line 2) to check if a PJH instance has already existed.  
When \emph{loadHeap} is finally invoked at line 4, the external name manager will locate the PJH instance and return its starting address by fetching the address hint.  
Afterwards, JVM will map the whole PJH at the starting address.
If the map phase fails due to the address occupied by the normal heap, we have to move the whole PJH into another virtual address. 
Since all the pointers within heap become trash, a thorough scan is warranted to update pointers. 
The remap phase might be very costly, but it may rarely happen thanks to the large virtual address space of 64-bit OSes. 
If the map operation succeeds, it will be followed by a class reinitialization phase.

The stock JVM will allocate a new Klass data structure in its metadata space for each class initialization. 
However, if we bluntly create new Klasses in the Klass segment during class reinitialization, all class pointers in PJH will become trash, which is unacceptable.
To avoid invalidating class pointers, we require that all Klasses in PJH stand for a place holder and be initialized in place. 
In this way, all objects and class pointers will become available after class reinitialization. 
Our design makes the load phase of PJH very fast because the time overhead is directly proportional to the number of Klasses instead of objects. 
Meanwhile, the number of Klasses in the Klass Segment is usually trivial. 
For example, a typical TPCC~\cite{tpcc} workload only requires nine different data classes to be persisted. 
%(discussed in~\ref{subsec:motiv})
After class reinitialization, \emph{loadHeap} will return and users are free to access the persistent data in the loaded PJH instance.

\textbf{Root-related APIs.} 
\emph{Root objects} marks some known locations of persistent objects and can be used as entry points to access PJH especially when a PJH instance is reloaded.  
\emph{getRoot} and \emph{setRoot} serve as getter/setter for the root objects. 
When \emph{getRoot} is called at line 5, the corresponding object \emph{p} will be returned. 
Since we don't store the type of the root object, the return type will be \emph{Object}, and users are responsible for type casting. 
After that, users can fetch other persistent data by accessing \emph{p}. 
Similarly, \emph{setRoot} at line 11 receives an object in arbitrary type and stores its address in the root table with the specified name for future use. 
%Root objects are useful especially when a PJH instance is reloaded. 
%Users need firstly invoke \emph{getRoot} to acquire a root object and get other objects by following its references. 

\subsection{Memory Safety}
The design of PJH has decoupled the persistence between an object and its fields: an object can be stored in NVM with a reference to DRAM. 
This design may cause a disaster if users try to access a pointer to volatile memory after a heap reloading. 
The pointer can point to anywhere and the modification of the referenced data can cause undefined consequences. 
In contrast, an over-restricted invariant on references is safe but at the sacrifice of usability.  
To this end, we have provided four different memory safety levels according to various requirements on usability and safety. 

\begin{itemize}
\item User-guaranteed safety. 
Users need to be aware of the presence of volatile pointers and avoid directly using them after a reload of PJH. 
This safety level lays the burden of checking on programmers and may cause unknown errors. 
However, it provides the best performance compared to others.

\item Zeroing safety. 
A PJH instance will first step into a check phase before loading, and all out pointers will be nullified. 
In this way, applications can easily tell if they have suffered a Java execution context loss with null-checks. 
Even the worst case for a careless access on invalid volatile pointers will only get a \emph{NullPointerException}, which is much better than one could experience in user-guarantee safety level. 

This level may be hard to implement in a heap design where no type-related information is kept, since integers may be mixed up with pointers and wrongly handled. 
However, it is achievable within PJH because the Klass segment has stored the required type information.  
The major disadvantage is that the check phase will traverse the whole heap and slow down the heap loading. 

\item Type-based safety.
For users who really want to access NVM safely, we have implemented a library atop Java to allow them defining classes with simple annotations, and only objects with those classes will be persisted into PJH (introduced in section~\ref{sec:jpo}). 
This safety level guarantees that no pointers within PJH will point out of it, and thus provides a similar safety level to NV-Heaps~\cite{coburn2011nv}. 
However, it requires that applications should be modified and annotated to fit the NVM. 

%\item Persist-By-GC safety.
%Type-based safety requires thorough control over the user-defined types, which may not be practical for all applications. 
%It also requires a third party library for the type transformation. 
%Hence, we also provide Persist-By-GC safety level, which leverages the garbage collection service in JVM to guarantee epoch-based persistence.  
%Persist-By-GC safety will ensure that all objects reachable from persistent roots in the name table are persisted after a garbage collection phase. 
%Since JVM maintains the information for types, it is capable to store objects in arbitrary types into PJH. 
%However, if the system crashes in the middle of an epoch with wrong pointers, PJH can only issue a warning and require either users (Users-guaranteed safety) or JVM (Zeroing safety) to check the heap. 
%Persist-By-GC could be useful for users who want to lazily persist some auxiliary generic data structures, such as a cache, for better reboot performance, with little programming effort. 
%We have leveraged Persist-By-GC safety to implement a simple NVM cache for Spark and the result will be shown in Section~\ref{sec:eval}. 

%This level is backed by our Persist-By-GC mechanism, which will be mentioned in detail in Section~\ref{sec:spark}. 
%Programmers can rely on the mechanism so that all non-volatile-to-volatile pointers will eventually be corrected. 
%However, if the system crashes in the middle of an epoch with wrong pointers, the PJH can only issue a warning and requires either users (Users-guaranteed safety) or JVM (Zeroing safety) to check the heap.  
\end{itemize}

\subsection{Persistence Guarantee}
Mainstream computer architectures only have volatile caches and thus require cache flush operations like \emph{clflush} to ensure data persisted in NVRAM. 
To preserve persistence ordering, we may further require memory fence instructions (\emph{sfence}).  
The \emph{pnew} keyword is only used for object allocation, so we can only provide persistence guarantee for heap-related metadata to build a recoverable heap to survive crashes (discussed in section~\ref{sec:crash}) with those instructions.
As for the application-level guarantee, we have provided some basic field-level APIs to manage the persistence of objects in a fine-grained way. 
%They are built atop Java Reflect APIs.
% since we are in an opinion that flush operations are "outsiders" to Java and should not be mixed with Java programming language.
Figure~\ref{fig:flush} illustrates an example to leverage our APIs. 
To persist a field in an object, we must fetch the incarnation of the field at runtime with Java Reflect APIs, such as \emph{getDeclaredField}. 
After that, we can use the newly added \emph{flush} interface to persist \emph{x.y}. 
If applications want to manipulate arrays, they can use \emph{Array.flush} to flush certain object with offset \emph{i}. 
The largest work set for those two APIs are restricted to 8 bytes to preserve atomicity. 
Besides, the implementation of those two APIs has added a \emph{sfence} instruction to ensure order.  
%The field-level API is very strict at the order of cache flush operations and sometimes may not be necessary according to user requirements. 

\begin{figure}[t]
{
\begin{q1}
Person x = pnew Person(...);
Person[] z = pnew Person[10];
// After some operations...
......
Field f = x.getClass.getDeclaredField("id");
// Newly added APIs below:
f.flush(x);        	    // for normal fields (flush x.y) 
Array.flush(z, 3);		// for arrays (flush z[3])
\end{q1}
}
\caption{A simple program to illustrate our flush APIs}
\label{fig:flush}
\end{figure} 

Additionally, we have also added a coarse-grained \emph{flush} method in the implementation of \emph{Object} class for performance consideration. 
This method will flush all the data fields in the object into NVM with only one \emph{sfence} in the end. 
It is suitable for scenarios where the persistent order among fields of an object doesn't matter.  
Other advanced features, such as transitively persist all data reachable from an object, can be easily implemented with those basic methods.

For users who want to manipulate their data in an ACID semantic, we have provided an abstraction called PJO, which will be discussed later.

\section{Crash-consistent Heap}
\label{sec:crash}
The design of PJH should consider crashes which can happen at any time to avoid inconsistency. 
To this end, {\sys} enhances the allocation and garbage collection phase to ensure that 
the heap can be recovered to a consistent state upon failure.

\iffalse
\subsection{Overview}  
Our PJH is implemented as an independent \emph{persistent space} in parallel with the original Java heap. 
The persistence space reuses some volatile data structures defined by \emph{young} and \emph{old} space to manage the persistent heap space. 
We will use \emph{persistent space} to denote our persistent heap in the rest of the section, in correspondence with the notions of \emph{young} and
\emph{old} spaces. 
The persistent space is arranged as a non-generational heap, and the garbage collection algorithm resembles the old GC in PSGC, in that it is designed 
for long-lived objects and infrequent collections. 
\fi

\subsection{Crash-consistent Allocation} 

The persistent heap maintains a variable named \emph{top} to memorize how much memory resource has been allocated. 
The value of \emph{top} is replicated in the PJH for future heap reloading. 
As we mentioned before, users are permitted to exploit \emph{pnew} to create a persistent object, which has an impact on the 
heap-related metadata. The allocation can be divided into three phases:

\begin{itemize}
 \item (1) Fetching the Klass pointer from the constant pool; 

\item (2) Allocating memory and updating the value of \emph{top}; 

\item (3) Initializing the object header. 
\end{itemize}

Since the Java compiler \emph{Javac} guarantees that an object will not become visible until it has finished initialization, {\sys} does not need to consider 
inferences with other threads.  
To make the allocation crash-consistent, the replica of the \emph{top} value in PJH should be persisted as soon as the modification on the volatile one in step (2), 
through cache flush and fence instructions. 
Otherwise, some created objects may be treated as unallocated and truncated during recovery due to the stale top value. 
Further, the Klass pointer update should be persisted in step (3) to avoid the situation where an initialized object refers to some corrupted Klass metadata.

\subsection{Crash-consistent Garbage Collection}

Since the life cycles for persistent objects are often long, we reuse the old GC algorithm in PSGC to collect them. 
However, the original algorithm is carefully enhanced for crash consistency. 

\paragraph{A Brief review of PSGC.}
PSGC exploits a three-phase region-based algorithm for its old GC. 
The whole heap has been divided into many small areas named \emph{regions}. 
The first marking phase will mark live objects from all roots. 
PSGC has maintained a read-only bitmap called \emph{mark bitmap} to memorize all live objects in a memory-efficient way. 

The second phase, namely \emph{summary phase}, will summarize the heap spaces and generate region-based indices to store the destination address of all live objects. 
After the summary phase, a region-to-region mapping will also be generated. 
Each region has a corresponding \emph{destination region} into which all live objects within it will be moved. 
Note that the summary phase is \emph{idempotent}: the indices and mappings are derived only from the mark bitmap, 
so the result of summary phase will be the same no matter how many times it executes, as long as the mark bitmap keeps intact. 

In the compact phase, the GC threads will pick out unprocessed regions and copy live objects into the destination regions. 
The regions will be processed concurrently; but each region will only be processed by one unique worker thread. 
For each object, the GC thread will first get its destination address by querying the region-based indices and copy its content. 
Afterwards, it will move to the copied object at the destination address, look into all references within it, and correct them respectively with the help of indices. 

\paragraph{Crash-consistent PSGC.}
Our algorithm extends the mark bitmap to mark live data objects in our persistent space during the marking phase. 
Since the mark bitmap can be seen as a sketch of the whole heap before the real collection, it must be persisted before the objects start being moved. 
After that, the heap will be marked as in the middle of a collection in the metadata area. 
The subsequent summary phase remains unchanged to generate region-based indices and the region-to-region mapping.

When the compact phase starts, the address of any objects could be changed and a crash happening in the middle might thereby 
corrupt the whole persistent heap. Therefore, {\sys} needs to make each step crash-consistent so that the persistent space can be recovered 
from failure and continue the collection. 

The object header has reserved several bits for PSGC; but it is only for young space and becomes useless once the object is promoted. 
{\sys} reuses those bits to implement a timestamp-based collection algorithm. 
Once a collection on the persistent space begins, {\sys} will update and persist the global timestamp in the metadata area within the persistent heap 
so that all objects in the space become stale. 
The timestamp of an object does not become valid until its whole content has been copied and persisted. 
If a crash happens in the middle of GC, we can tell whether objects are processed by simply inspecting the timestamp. 

Assume that we have known the destination address after querying the region-based indices. 
The copy algorithm contains three steps: 1). The first step is to copy the object directly to the destination address without modification;
2). Afterwards, {\sys} will start modifying the pointers in the copied object respectively. The data stored in the original address actually 
serves as undo log and can be leveraged for recovery; 3). Finally, {\sys} will modify the timestamp in both headers to the current global 
timestamp (but the copied one should be persisted first).  After setting the timestamp, the object will be treated as a processed object, 
and its data will remain intact regardless of any crashes. 

%Some special case may disable our method when the copied object at the destination address is overlapped with that at the source address. 
%In this case, we will install a counter on the header for the copied object using other metadata bits. 
%If those bits are being used, we will store them in a per-region 8-byte area in DRAM to make space for the counter. 
%As we mentioned before, those bits are only bound to a specific execution context, so it is harmless to abandon them after a crash.
%The counter will memorize how many pointers we have modified for an object. 
%If a crash happens during pointer modifications, we are able to find the first pointer we haven't modified and continue from there. 
%After all pointers are processed, the displaced bits will be reinstalled to the header, and the timestamp will be modified. 

Unfortunately, the original data serving as the undo log will not stay long. 
Once all live objects within a region have been evacuated, the region will be treated as reusable and soon becomes a destination region for other ones to overwrite. 
If {\sys} does not memorize which region has finished the processing, it will not be able to differentiate between a destination region 
which is half-overwritten and a source region which is half-copied after a crash. 
To address this issue, {\sys} maintains a \emph{region bitmap} in the metadata area to mark which region has been processed. 

\subsection{Recovery} 
The recovery phase will be activated by the API \emph{loadHeap} if the heap is marked as being garbage collected in the metadata area. 
The recovery also contains three steps: 1). The first task is to fetch the mark bitmap, the result of previous marking phase;
2). Afterwards, the summary phase will be redone by regenerating the volatile auxiliary data structure from the mark bitmap;
3). The last step is to fetch the region bitmap to locate the unprocessed or half-processed regions and process the objects within 
them using the same algorithm in the compact phase.  
After recovery and the \emph{loadHeap} returns, the whole heap can be safely used by applications. 

\section{Persistent Java Object}
\label{sec:jpo}

PJH only guarantees crash consistency for heap-related metadata; application data may still be corrupted upon a crash. 
While a \emph{persistent layer} like JPA is helpful to provide a convenient persistent programming model, it incurs high overhead
upon NVM. 
To this end, {\sys} builds persistent Java object (PJO) atop the persistent Java heap (PJH) as an alternative for persistent programming. 

PJH already allows applications like databases to store their data in NVM as a normal Java object. This offers opportunities to rethink about the
persistent layer. 
PJO provides backward compatibility through reusing the interfaces and annotations
provided by JPA. Yet,  it reaps the benefits offered by PJH with much better performance. 

\begin{figure}[h]
\begin{minipage}{.99\linewidth}
\centering
\includegraphics[scale=0.35]{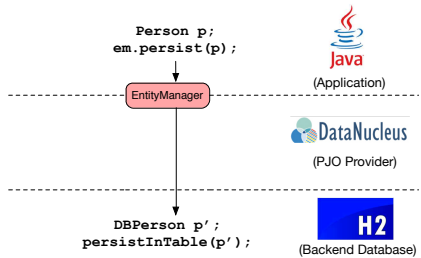}
\caption{NVM-aware infrastructure of DataNucleus}
\label{fig:datanucleusnew}
\end{minipage} \\[-10pt]
\end{figure}

Figure~\ref{fig:datanucleusnew} illustrates the modified architecture of database frameworks with PJO. 
The programmer can still use \emph{em.persist(p)} to persist a \emph{Person} object into NVM. 
However, when real persistent work begins, data in \emph{p} will be directly shipped to the backend database. 
The PJO provider still helps manage the persistent objects, but the SQL transformation phase is removed. 

{\sys} provides a new lightweight abstraction called \emph{DBPersistable} to support all objects actually stored in NVM. 
The \emph{DBPerson} class in Figure~\ref{fig:datanucleusnew} is an example of \emph{DBPersistable}. 
A \emph{DBPersistable} object resembles the \emph{Persistable} one except that the control fields related to PJO providers are stripped.

\begin{figure}[h]
\begin{minipage}{.99\linewidth}
\centering
\begin{tabular}{cc}
\subfloat[An enhanced \emph{Person} object stored in DRAM] {
\includegraphics[scale=0.24]{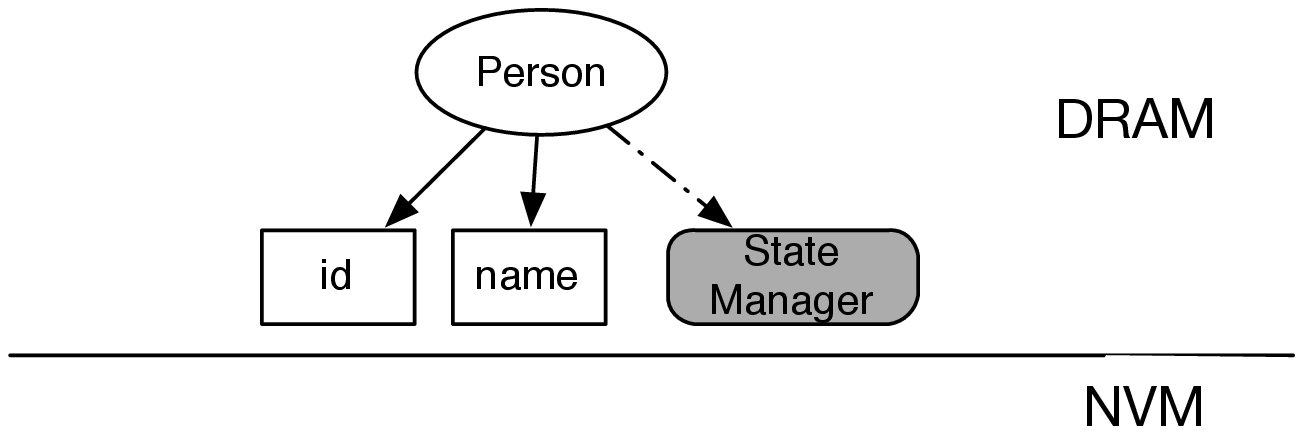}
\label{subfig:objlayout0}
}
%\hspace{0.01\textwidth}
& \subfloat[Creating a \emph{DBPerson} object during persisting] {
\includegraphics[scale=0.24]{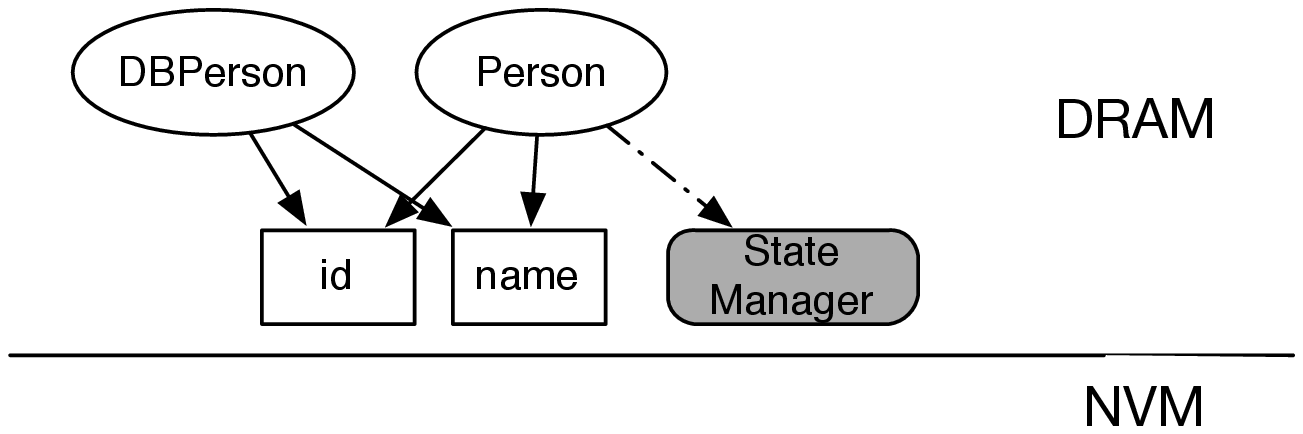}
\label{subfig:objlayout1}
}
\\
\vspace{0.01\textwidth}
\subfloat[Persisting the \emph{DBPerson} object] {
\includegraphics[scale=0.24]{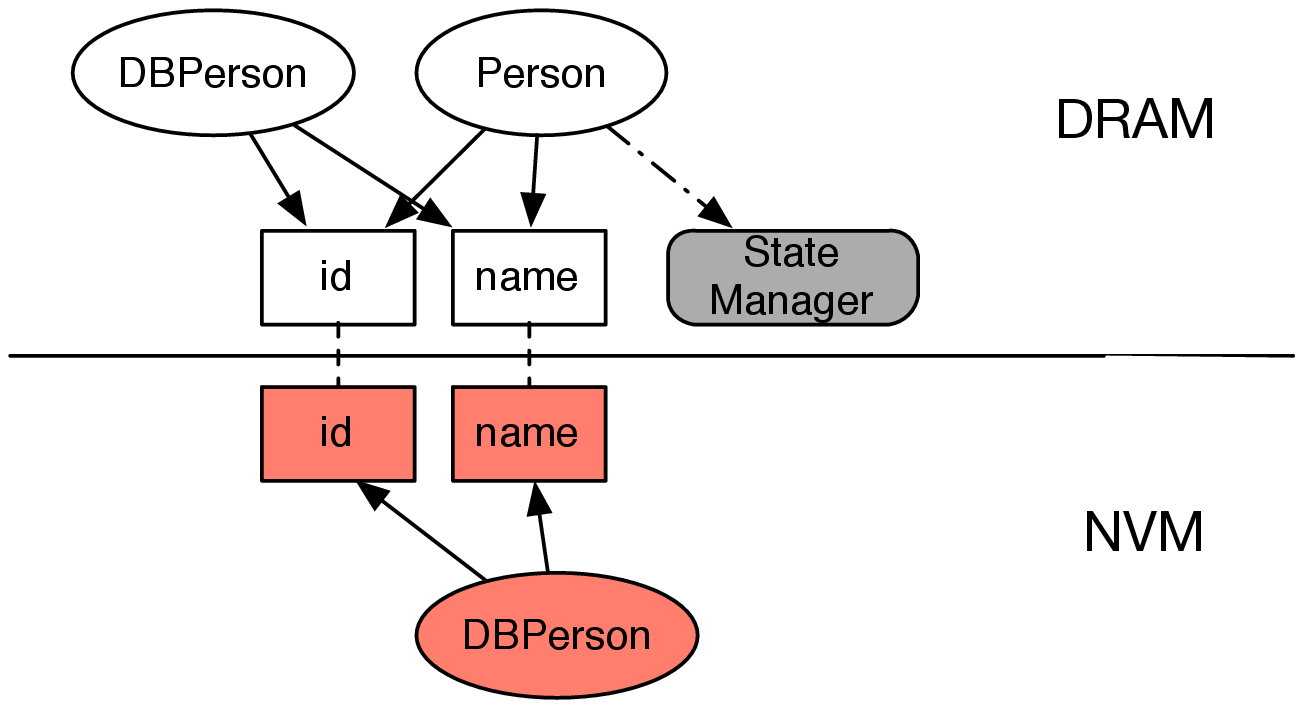}
\label{subfig:objlayout2}
}

%\hspace{0.01\textwidth}
& \subfloat[Data deduplication for the original \emph{Person} object] {
\includegraphics[scale=0.24]{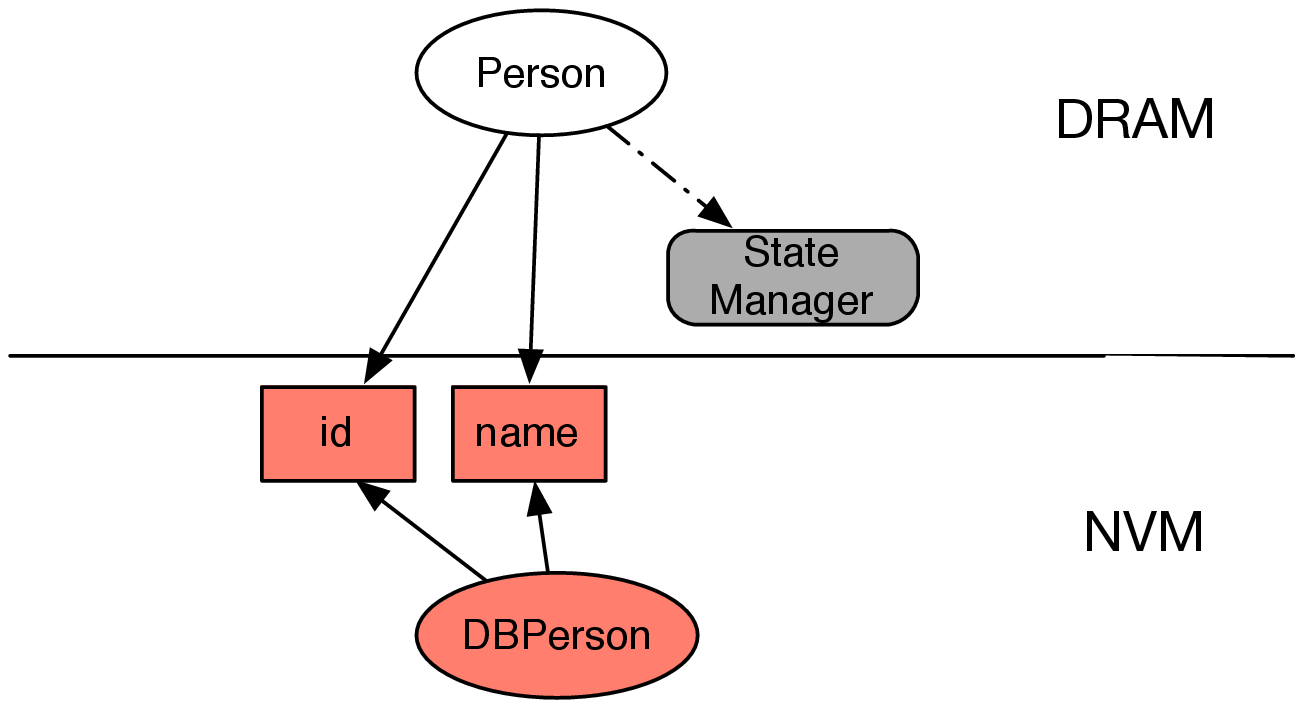}
\label{subfig:objlayout3}
}
\\
\vspace{0.01\textwidth}
\\[-16pt]
\end{tabular}
\caption{A detailed example to show how PJO exactly works.}
\label{fig:objlayout}
\end{minipage}
\end{figure}

Figure~\ref{fig:objlayout} shows how PJO exactly works for a persist operation on a \emph{Person} object, 
whose data fields (\emph{id} and \emph{name}) are referenced by solid lines.
The PJO provider (our modified \emph{DataNucleus}) will enhance \emph{Person} so that each object keeps a \emph{StateManager} field for 
metadata management and access control (referenced by a dash line). 
The \emph{StateManager} field is transparent with applications.  
When persisting, a corresponding \emph{DBPerson} object will be generated with all its data fields referenced to the \emph{Person} object (Figure~\ref{subfig:objlayout1}). 
The \emph{DBPerson} object will be shipped to the backend database for data persistence. 
The most straightforward implementation is to directly persist it into NVM as illustrated in Figure~\ref{subfig:objlayout2}; 
but the backend database is free to conduct any operations on the object, such as logging. 
This provides flexibility for database frameworks to implement their own ACID protocols. 

We have implemented a PJO provider by modifying \emph{DataNucleus}. 
It provides the same APIs as JPA such that no modification to applications is required. 
Programmers can leverage the APIs provided by the PJO provider to retrieve, process and update data in an object-oriented fashion. 
Furthermore, we have also implemented some advanced features to make PJO more productive. 

\textbf{Data deduplication.} 
Once the data objects are persisted, the volatile copy left in DRAM becomes redundant. 
However, the memory resource cannot be reclaimed because the user-defined objects still hold references to them. 
We have implemented a data deduplication optimization within the DataNucleus enhancer such that the data fields of objects will 
be redirected to the persistent data after a transaction commits.
As illustrated in Figure~\ref{subfig:objlayout3}, all data fields in the original \emph{Person} object has been modified to point to the persisted data. 
Consequently, pervious volatile fields can be reclaimed or reused to save memory. 

\textbf{Field-level tracking.} 
Since persistent data now is arranged as Java objects, PJO enables field-level manipulation by tracking modified fields in a transaction. 
The enhanced Person objects will maintain a bitmap to mark if its fields have been modified respectively. 
During commit, the bitmap will be sent to the backend database together with the DBPerson object. 
The backend database can thereby only update the modified fields in the persistent storage. 
This also helps to reduce logging overhead if the backend database has implemented a field-level logging. 

Field-level tracking is also useful when data deduplication is enabled. 
When the field should be updated, a copy-on-write phase will happen, and a shadow, non-persistent field will be created for future write. 
The modified non-persistent field will be flushed into backend database during commit. 
Such a design is motivated by the fact that write latency in emerging NVM will be several times larger than DRAM while read latency rivals DRAM~\cite{Xia2017HiKV}. 
More importantly, it avoids careless (or malicious) update without protection to corrupt the persistent storage. 

\textbf{Abundant types.} 
Similarly to JPA, our PJO has also supported various types for end users. 
Users are free to use inherited classes, collections and foreign-key-like references to build their own database applications; 
the performance results will be illustrated in section~\ref{sec:eval}. 

We currently mainly use PJO to support database applications. 
Yet, it can be further extended into a general framework to provide an easy-to-use 
persistent programming model. With its transaction APIs and bytecode instrumentation tools (enhancer), applications are free to manipulate 
their objects and expect they are persisted in an ACID fashion.

\section{Evaluation}
\label{sec:eval}

%\textcolor{red}{The evaluation will be arranged below:
%1. micro benchmark: comparison against PCJ; heap loading time.
%1. database: test over JPAB. There are 4 testcases within JPAB and all of them will test CRUD operations. There will be four figures as a group.  
%2. Spark: Overall execution time and recovery time in a 1.4G dataset (no figure, in words), the job execution time distribution graph to show the improvement (1 figure), breakdown analysis (1 figure). 
%3. heap load time (no figure, in words) 
%}

\subsection{Experiment setup}
We have implemented PJH on OpenJDK 8u102-b14, which comprises approximately 7,000 lines of C++ code and 300 lines of Java code. 
We have also modified DataNucleus to implement PJO with 1,500 lines of Java code. 
As for the backend database H2, it takes about 600 LoC to make it support both PJO (mainly for the \emph{DBPersistable} interface) 
and PJH (mainly replacing \emph{new} with \emph{pnew}). 
The data structures for transaction control (like logging) remain intact. 
The modification is minor considering the whole code base of H2 (about 14K LoC). 
In contrast, a design like PCJ would require a thorough rewrite over the main data structures in H2 to fit NVM. 

Our evaluation is conducted on a machine with dual Intel \textregistered Xeon\textsuperscript{TM} E5-2618L v3 CPUs (16 cores).  
It contains 64G DRAM and 64G Viking NVDIMM device. 
The operating system is Linux-4.9.6. 
We set the maximum Java heap size to 8G for evaluation. 

\subsection{Comparison with PCJ}
\begin{figure}[h]
\begin{minipage}{.99\linewidth}
\centering
\includegraphics[scale=.6]{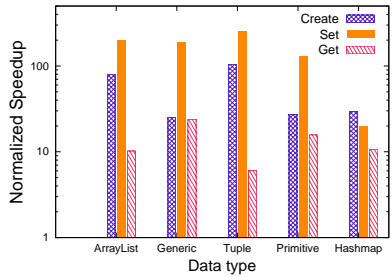}
\caption{Normalized speedup for PJH compared to PCJ}
\label{fig:pcj}
\end{minipage}\\[-4pt]
\end{figure}

PCJ provides an independent type system against the original one in Java including tuples, generic arrays and hashmaps. 
We also implement similar data structures atop our PJH. Since PCJ provides ACID semantics for all operations, 
we also add ACID guarantee by providing a simple undo log to make a fair comparison. 
The microbenchmarks conduct millions of primitive operations (create/get/set) on those data types and then collect the execution time.  
The results are shown in Figure~\ref{fig:pcj}. 

Surprisingly, our PJH greatly outperforms against PCJ, and the best speedup even reaches 256.3x for set operations on tuples. 
PJH performs much better in \emph{set} and \emph{create} operations in that PCJ stores data off heap and thus require a complicated metadata update for those operations. 
As for \emph{get} operations, the improvement of PCJ drops due to less requirement for metadata management, but it still outperforms PCJ by at least 6.0x.

\vspace{5mm}

\begin{table}[h]
\fontsize{7.5}{9.6}\selectfont
\centering
\begin{tabular}{ll}
	\hline
	Name & Description \\
	\hline
	BasicTest & Testing over basic user-defined classes\\
	
	ExtTest & Testing over classes with inheritance relationships\\
	
	CollectionTest & Testing over classes containing collection members\\
	
	NodeTest & Testing over classes with foreign-key-like references\\
	\hline
\end{tabular}
\\[-5pt]
\caption{The description for each test cases in JPAB}
\label{tab:jpab}
\end{table}

\subsection{Comparison with JPA}

\begin{figure*}[t]
\noindent
\centering
\subfloat[BasicTest] {
\includegraphics[scale=0.48]{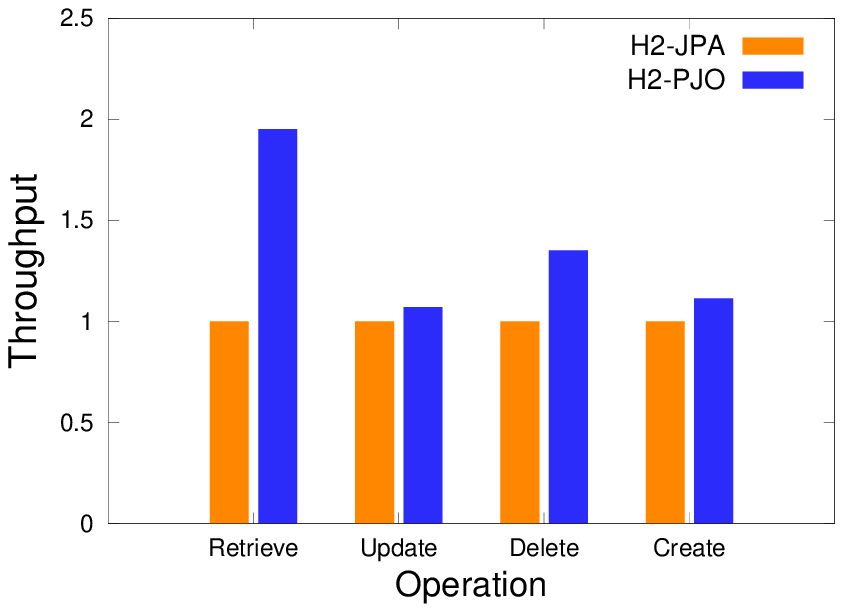}
\label{subfig:BasicTest}
}
%\hspace{0.02\textwidth}
\subfloat[ExtTest] {
\includegraphics[scale=0.48]{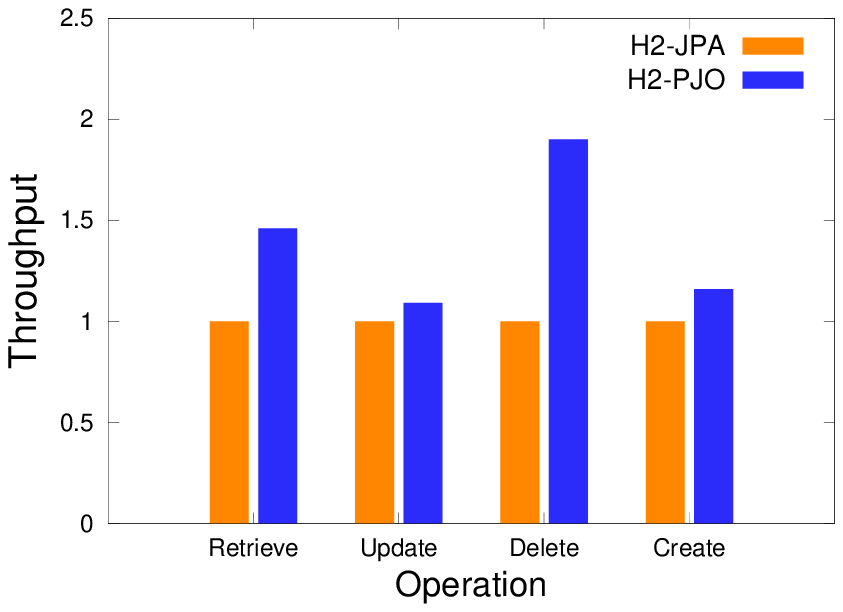}
\label{subfig:ExtTest}
}
\subfloat[CollectionTest] {
\includegraphics[scale=0.48]{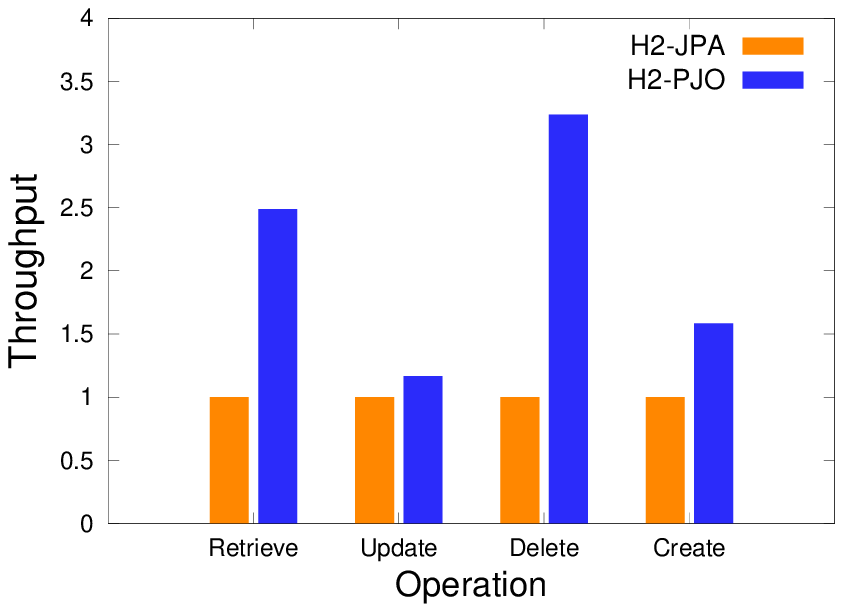}
\label{subfig:CollectionTest}
}
%\hspace{0.02\textwidth}
\subfloat[NodeTest] {
\includegraphics[scale=0.48]{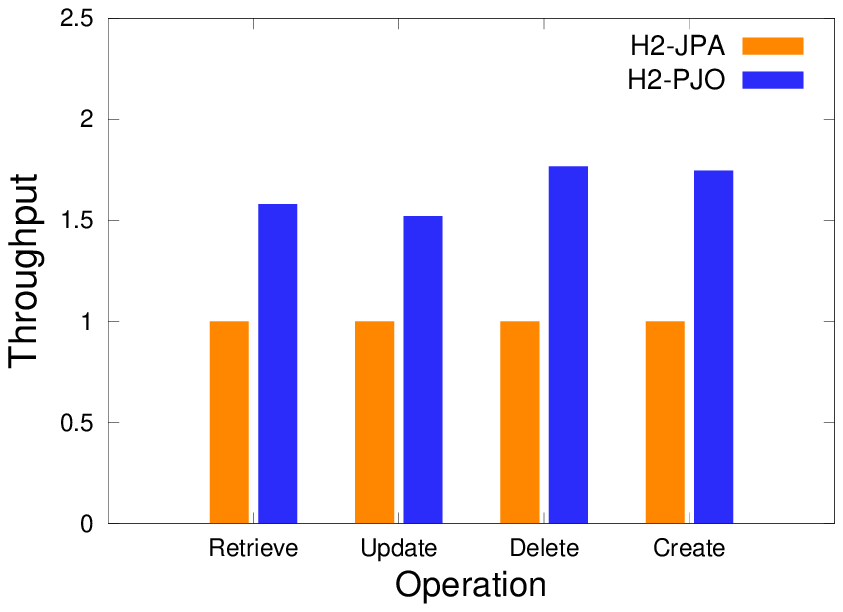}
\label{subfig:NodeTest}
}
\\[-5pt]
\vspace{0.01\textwidth}
\caption{Evaluation for JPAB benchmark}
\label{fig:eval-jpab}
\end{figure*}

We use the JPA Benchmark (JPAB)~\cite{JPAB} to compare PJO and JPA, whose detailed description is illustrated in Table~\ref{tab:jpab}.
JPAB contains normal CRUD\footnote{CRUD means four basic operations of persistent storage: create, read, update and delete} operations 
and tests over various features of a JPA framework, such as inheritance, collections and foreign keys. 
We use unmodified JPA and H2 running on NVDIMM as for the baseline (H2-JPA).
The evaluation result in Figure~\ref{fig:eval-jpab} indicates that PJO (H2-PJO) outperforms H2-JPA in all test cases and provides up to 3.24x speedup. 

\begin{figure}[h]
\begin{minipage}{.99\linewidth}
\centering
\includegraphics[scale=.6]{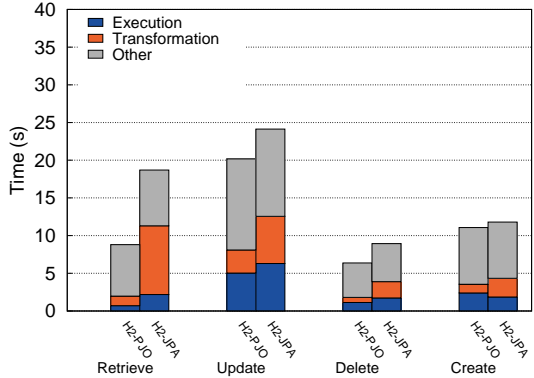}
\vspace{0.01\textwidth}
\caption{Breakdown analysis for BasicTest}
\label{fig:breakdown-pjo1}
\end{minipage} \\[-10pt]
\end{figure}

We have also exploited \emph{BasicTest} as an example to provide a detailed analysis. 
We break down the performance into three parts: execution in H2 database, transformation for SQL statements and others. 
As illustrated in Figure~\ref{fig:breakdown-pjo1}, the transformation overhead is significantly reduced thanks to PJO. 
Furthermore, the execution time in H2 also decreases for most cases, which can be attributed to the interface change from the
JDBC interfaces to our \emph{DBPersistable} abstractions. 

%The improvement is mainly related to the complexity to generate, transform and compile SQL statements. 
%For example, the most speedup comes from the remove and retrieve operations for collections, since the SQL statements generated by original DataNucleus are rather complex to process. 
%In summary, our evaluation confirms that our JPO can provide better performance while preserving easy-to-use interfaces for Java end users. 

%The improvement is limited for retrieve operations, in that the original DataNucleus maintains an object cache to store the recently accessed persistent data within JVM to boost the performance. 
%If the cache is hit, the object will be returned to users and no SQL transformation is required, so the throughput is close to our JPO which also put objects in memory. 

%\textcolor{red}{TODO: provide more explanation on the data. }

\subsection{Microbenchmark}

\begin{figure}[h]
\begin{minipage}{.99\linewidth}
\centering
\includegraphics[scale=.6]{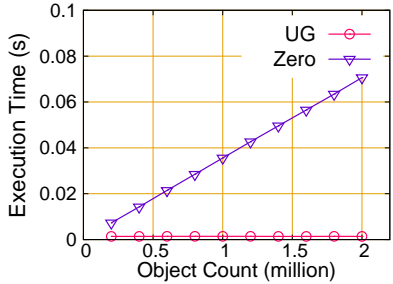}
\caption{Heap loading time with user-guaranteed (UG) and zeroing (Zero) safety.}
\label{fig:heap-load}
\end{minipage}\\[-8pt]
\end{figure}

\textbf{Heap loading time. }
We test the heap loading time with a micro-benchmark which generate a large number of objects (from 0.2 million to 2 million) of 20 different Klasses. 
Besides, we evaluate heap loading with both user-guaranteed (UG) and zeroing (Zero) safety. 
As shown in Figure~\ref{fig:heap-load}, the heap loading time for user-guaranteed safety remains constant when the number of objects increases, as the heap loading is dominated by the number of Klasses instead of objects. 
In contrast, the loading time grows linearly with the number of objects with zeroing safety since it requires a whole heap scan to validate all objects. 
When the number of objects reaches 2 million, the heap loading time is about 72.76ms, which is still trivial compared to the JVM warm-up time studied by previous work~\cite{lion2016don}. 

\textbf{The cost of recoverable GC.}
We use a micro-benchmark to test the overhead of our recoverable garbage collection mechanism. 
The benchmark allocates lots of objects (about 1GB) on PJH and some references to them are abandoned afterwards. 
We use \emph{System.gc()} to forcedly collect PJH and test the pause time. 
For the baseline, we remove all the clflush operations to make the algorithm nearly the same as the original old GC. 
The evaluation result shows that the flush operations would increase the pause time by 17.8\%, which is still acceptable
for the benefit of crash consistency.

%\textbf{Persist-by-GC effects.} 
%We have implemented a simple \emph{MemoryPersister} class to allow Spark to cache RDDs in PJH lazily. 
%Since Spark always generates a unique id for each application, we use it as the name for PJH instances. 
%Once an executor crashes and restarts, the Spark runtime will try to find if a PJH instance whose name matches current application id. 
%If so, the PJH instance will be loaded and all RDDs within it will be marked as usable. 
%The overall modification on Spark only takes about 150 LOCs thanks to the Persist-By-GC mechanism. 
%
%We evaluate Spark over \emph{JavaHdfsLR}, a benchmark contained in the original Spark paper~\cite{zaharia2010spark}. 
%The size of dataset is around 1.6 GB. 
%The baseline uses the \emph{checkpoint} interface to store RDD in hdfs for recovery. 
%During execution, we kill the executor process and clean the file cache to simulate a machine failure. 
%The result shows that we can reduce the recovery time by 66.7\%. 

%\input{results}

%\input{challenge}

\section{Related Work}
\label{sec:related}

\paragraph{Non-volatile Memory Heaps} 
The invention of recoverable VM~\cite{satyanarayanan1994lightweight} once stimulates research on building persistent, recoverable and efficient heaps for user-defined objects~\cite{lowell1997free, o1994concurrent}. 
Orthogonally persistent Java~\cite{atkinson1996orthogonally, jordan2001early, jordan1999orthogonal} is proposed to provide a whole-system-persistent Java runtime, 
including a non-volatile heap.  However, it has to cope with tricky stuffs like the \emph{System} class. 
Subsequent work turns to persistent object stores~\cite{liskov1996safe, paterson2006db4o, white1994quickstore} and entity-relation mappings~\cite{demichiel2006java, jordan2003java} for practice. 
{\sys} instead discards the requirement of whole-system persistence and provides both coarse-grained and fine-grained persistence atop NVM.

The topic on persistent memory heaps has been renewed recently due to the development of NVM technology. 
NV-Heaps~\cite{coburn2011nv} pioneers in specifying cross-heap pointers. 
They avoid potential memory leaks by directly disabling non-volatile-to-volatile pointers with a compiler-time checker.
However, this restriction precludes scenarios where applications leverage both DRAM and NVM, which are common in state-of-art NVM-based systems. 
Besides, NV-Heaps provides a simple reference-counting garbage collector, which is shown to be ineffective~\cite{shahriyar2013taking}. 
Our PJH supports flexible pointers and integrates while reusing the garbage collector in JVM. 
Makalu~\cite{bhandari2016makalu} is a persistent memory allocator built on the programming model of Atlas~\cite{chakrabarti2014atlas}. 
It provides persistent and recoverable guarantee for the allocation metadata and leverages a recovery-time garbage collector. 
{\sys} also considers the crash consistency for the heap but the garbage collection is online thanks to Java's GC service.

\paragraph{Java Runtime Optimization}
Improving the efficiency of Java runtime has drawn large attention due to its wide utilization in large-scale applications nowadays.  
HotTub~\cite{lion2016don} finds that class loading is an important source of inefficiency during JVM warm-up and introduces a pool with virtual machines 
whose classes have been loaded to mitigate the overhead. 
OpenJDK also brings in \emph{Class Data Sharing} (CDS) to allow multiple JVMs to share precompiled classes for fast startup. 
However, this feature is still experimental for the moment. 
Our work shares similar wisdom of reducing loading time but in a different way through NVM. 
In PJH, we provide the Klass segment which stores some placeholders for Klasses. 
During reloading, Klasses will be reinitialized and stored in place so that all class pointers within persistent objects remain valid. 

%AOT (Ahead-Of-Time) compilation is proposed as an alternative way for JIT. 
%In AOT, the early-compiled binary will be persisted so as to be reused by multiple JVMs, which is also useful to reduce warm-up time. 
%Maas et al.~\cite{maas2017return} argues that AOT would be efficient in a cloud environment by implementing a code cache which can survive executions, which suits the design of PJH. 
%An extension of PJH can include a non-volatile code cache for further reuse. 
%Since the compiled code is almost write-once but read many times, it is also suitable to store in NVM. 
 
Another line of work studied the performance of garbage collectors and leverage different ways to optimize them. 
NumaGiC~\cite{gidra2013study, gidra2015numagic} finds that the old garbage collector in Java suffers from scalability issues due to NUMA-unawareness and solves the problem with a NUMA-friendly algorithm. 
Yu et al.~\cite{yu2016performance} spot out a performance bottleneck during new pointer calculation which can be resolved with caching previous results. 
Cutler et al.~\cite{cutler2016reducing} try to avoid unnecessary object copying and compacting in full garbage collector with clustered analysis while 
Yak GC~\cite{nguyen2016yak} suggests a region-based algorithm. 
These are orthogonal to ours and some would be helpful to optimize our recoverable GC. 

\paragraph{Other NVM-based Systems}
Transactions are a hot topic in building NVM-backed systems. 
Mnemosyne~\cite{volos2011mnemosyne} implements semantic-free raw word log (RAWL) in support of transactions. 
Atlas~\cite{boehm2016persistence, chakrabarti2014atlas} instead uses synchronization variables like locks and recovery code to provide transaction-like ACID properties, and NVThreads~\cite{hsu2017nvthreads} tries to optimize it with a coarse-grained logging protocol.  
Other work~\cite{Dong2017Soft, kolli2016high, liu2017dudetm, memaripour2017atomic} points out that the persist operations (including clflush) should not be included in the critical path of transactions and provide various solutions to move them background. 
We also provide an abstraction named PJO to provide transaction interfaces.  

Other systems combine other other emerging hardware features with NVM. 
Octopus~\cite{lu2017octopus} leverages RDMA technology together with NVM to build a distributed persistent memory file system.
Mojim~\cite{zhang2015mojim} also exploits RDMA and NVM, but aims at large scale storage systems where reliability and availability are critical. 
Seo et al.~\cite{seo2017failure} employs Intel's Restricted Transactional Memory to build an NVM-aware slotted-page structure for databases. 
{\sys} could also benefit from them once they are introduced into the Java world.

\section{Conclusions}
\label{sec:concl}

This paper proposed \emph{Espresso} to enable Java programmers to exploit NVM to ease persistence management. 
{\sys} comprised Persistent Java Heap (PJH) and Persistent Java Object (PJO) atop PJH.  
Evaluation showed that the eased persistence management resulted in notable performance boost.

\balance

\sloppy
{\footnotesize 
\bibliographystyle{abbrv}
\bibliography{nvm}}

\end{document}